\documentstyle[preprint,aps,eqsecnum,epsf,floats]{revtex}
\tighten

\begin{document}

\preprint{\vbox{\hbox{UTPT--95-11}
\hbox{CMU--HEP95--06}
\hbox{DOE--ER/40682--96}
\hbox{JHU--TIPAC--950007}
\hbox{hep-ph/9507284}}}

\title{Hadron Spectra for Semileptonic Heavy Quark Decay}

\author{Adam F.~Falk}
\address{Department of Physics and Astronomy, The Johns Hopkins
University\\ 3400 North
Charles Street, Baltimore, Maryland 21218 U.S.A.\\ {\tt
falk@planck.pha.jhu.edu}}
\author{Michael Luke}
\address{Department of Physics, University of Toronto\\60
St.~George
Street, Toronto, Ontario,
Canada M5S 1A7\\ {\tt luke@medb.physics.utoronto.ca}}
\author{Martin J. Savage}
\address{Department of Physics, Carnegie Mellon University\\
Pittsburgh, Pennsylvania 15213
U.S.A.\\ {\tt savage@thepub.phys.cmu.edu}}

\bigskip
\date{July 1995}

\maketitle
\begin{abstract} We calculate the leading perturbative and
power corrections to the hadronic invariant mass
and energy spectra in semileptonic heavy hadron decays.
We apply our results to the $B$ system.
Moments of the invariant
mass spectrum, which vanish in the parton model, probe gluon
bremsstrahlung and nonperturbative effects.  Combining our results with
recent
data on $B$ meson branching ratios, we obtain a lower bound
$\bar\Lambda>410\,{\rm MeV}$ and an upper bound $m_b^{\rm
pole}<4.89\,$GeV. The
Brodsky-Lepage-Mackenzie scale setting procedure suggests that higher order
perturbative
corrections to the first moment of the hadronic invariant mass spectrum are
small for bottom decay, and even tractable for charm decay.

\end{abstract}

\pacs{13.20.He, 12.38.Bx, 13.20.Fc, 13.30.Ce}

\section{Introduction}

Our understanding of inclusive decays of hadrons containing at least one
heavy quark has improved greatly over the last few years.  The energy
released  during semileptonic or radiative  decay of heavy hadrons is much
larger than the scale $\Lambda_{\rm QCD}$ of the strong  interactions,
and therefore an operator product expansion (OPE) exists for
some  observables in these decays,  including rates and differential
spectra \cite{SV,CGG}. The leading power corrections to the rates and
lepton differential spectra for semileptonic decays of heavy
hadrons\cite{Bigietc,MW,Mannel,FLNN} have been studied extensively, as have
the power corrections to radiative decays \cite{rare}.

A major result of this analysis is that, except in regions where the
expansion becomes
singular such as the endpoint of the electron spectrum in semileptonic
$b\rightarrow
u$ decay, the corrections to the parton model are quite small, suppressed
by
${\cal O}(\Lambda_{\rm QCD}^2/m_b^2)$.  While this does mean that the
parton model
is quite successful, it makes it difficult to test quantitatively the
corrections given by the
OPE.  In particular, if quark/hadron duality does not hold in this energy
regime, one would expect to see corrections to the parton model which
could not be accounted for by the leading perturbative and $1/m_Q$
corrections.
Shifman has recently criticized the related
OPE analysis of $\tau$ decay on the basis that violations of
duality in the Minkowski regime
introduce large corrections which are not seen at any finite order
in the OPE~\cite{Shifman}.

In this paper we suggest that hadronic variables, in particular moments of
the
invariant mass spectrum ${\rm d}\Gamma/{\rm d}s_H$ and the hadron energy
spectrum ${\rm d}\Gamma/{\rm d}E_H$, provide a
useful testing ground for the OPE.  This is similar to the analogous
suggestion, and analysis, for semileptonic $\tau$
decays~\cite{DP,CLEO}.  However, unlike the case for $\tau$ decays,
at tree level the final hadronic state at the parton level consists
of a single quark.  Therefore at lowest order in the OPE the final
hadronic state has fixed invariant mass $s_H=m_q^2$, and positive moments
of
$(s_H-m_q^2)$, which are calculable as a double expansion in
$\alpha_s(m_b)$
and
$1/m_b$, directly probe
physics beyond the parton model. Similarly, at leading order in the OPE
the maximum hadron energy is
$(m_b^2+m_q^2)/2 m_b$ (when the quark $q$ recoils back-to-back with the
leptons); the region above this endpoint is populated only by gluon
bremsstrahlung and nonperturbative effects.

In this paper we calculate the
corrections to the parton model results for these observables, up to
${\cal O}(1/m_b^2,\alpha_s/m_b)$.   As discussed in Ref.~\cite{BU},
although
the leading power corrections to leptonic variables arise at ${\cal
O}(1/m_b^2)$, for kinematic reasons the leading power
corrections to moments of the invariant mass spectrum arise at ${\cal
O}(1/m_b)$.  The ${\cal O}(1/m_b^2)$
corrections to the differential hadronic energy spectrum were
first examined in Ref.~\cite{BZ}; however, we disagree with the results
presented in
that work.   We also use the results of Ref.~\cite{CJK}, in which the
one-loop
corrections to the hadron energy spectrum were calculated.  We combine our
results with recent data on $B$ meson branching ratios to obtain a lower
bound
on the nonperturbative parameter $\bar\Lambda$, which is the leading
contribution to the difference between heavy quark and heavy meson masses.

Finally, using the BLM prescription~\cite{BLM} to estimate the size of
the two-loop perturbative corrections to the moments of the invariant
mass spectrum, we demonstrate that the first moment appears to have a
well-behaved
perturbative expansion not only for $B$ decays, but also for $D$ decays,
when the results are expressed in
terms of physical observables.   This suggests that studying hadronic
observables in semileptonic decays of charmed hadrons, which are dominated
by
only two or three resonances, may shed some insight into the applicability
of
quark/hadron duality at low energies.

\section{Kinematics}

We start by introducing the kinematic variables describing the
final state hadrons.  For definiteness, we will consider
semileptonic $B$ decay, although the analysis extends simply to the
decays of charmed hadrons.

\begin{figure}
\epsfxsize=10cm
\hfil\epsfbox{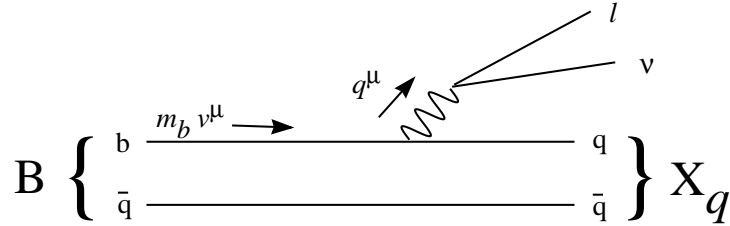}\hfill
\caption{The kinematics for $B\to X_q\ell\nu$.}
\label{kinematics}
\end{figure}

The kinematics of the inclusive process $B\to X_q\ell\nu$ is shown in
Fig.~\ref{kinematics}.
The four-momentum of the $B$ meson is $P_B^\mu=m_Bv^\mu$, and
$q^\mu$ is the  four-momentum of the lepton pair.  We write the
four-momentum
of the $b$ {\it quark\/} as $P_b^\mu$, and assign the heavy quark the same
four-velocity $v^\mu$ as the heavy meson.  The total energy of the leptons
in
the $B$ rest frame is $v\cdot q$, and their invariant mass is $q^2$.  It is
convenient to define dimensionless parton level quantities $\hat E_0$  and
$\hat s_0$,
\begin{eqnarray}\label{relations}
  \hat E_0&=& v\cdot (P_b - q)/m_b  = 1-v\cdot\hat q\,, \\
  \hat s_0&=& (P_b-q)^2/m_b^2  =  1-2v\cdot\hat q +\hat q^2\,,
\nonumber
\end{eqnarray}
where $\hat q^\mu=q^\mu/m_b$.
At leading order in $1/m_b$, $\hat E_0$ and $\hat s_0$ are simply  the
scaled energy and squared invariant mass of the final hadronic state.
However, since they are scaled by the $b$ {\it quark} mass,  this
identification does not hold at subleading order in
$1/m_b$.\footnote{This fact was neglected in Ref.~\cite{BZ}.}   Instead,
they are related to the physical hadronic energy and  squared   invariant
mass,
\begin{eqnarray}
  E_H &=& v\cdot (P_B - q)  =  m_B-v\cdot q\,, \\
  s_H&=& (P_B-q)^2  =  m_B^2-2m_Bv\cdot q+q^2\,,  \nonumber
\end{eqnarray}
through
\begin{eqnarray}\label{physparton}
  E_H&=&\bar\Lambda-{\lambda_1+3\lambda_2\over2m_B}+\left(m_B-\bar
  \Lambda+
  {\lambda_1+3\lambda_2\over2m_B}\right)\hat E_0+\dots\,,\nonumber\\
  s_H&=&m_q^2+\bar\Lambda^2+(m_B^2-2\bar\Lambda m_B+\bar\Lambda^2
  +\lambda_1+3\lambda_2)\,(\hat s_0-\hat m_q^2)\\
  &&\qquad\qquad\mbox{}+(2\bar\Lambda
m_B-2\Lambda^2-\lambda_1-3\lambda_2)\hat
E_0
  +\dots\,,\nonumber
\end{eqnarray}
where $\hat m_q=m_q/m_b$, and the ellipses denote terms higher order in
$1/m_b$.
The quantities $\bar\Lambda$, $\lambda_1$ and $\lambda_2$ arise in the
relationship between the quark and meson masses~\cite{Luke,FN},
\begin{eqnarray}\label{massrelate}
  m_B&=&m_b+\bar\Lambda-{\lambda_1+3\lambda_2\over2m_b}+\dots\,,\nonumber\\
  m_{B^*}&=&m_b+\bar\Lambda-{\lambda_1-\lambda_2\over2m_b}+\dots\,.
\end{eqnarray}
 From the measured $B$--$B^*$ mass splitting, $\lambda_2\simeq 0.12\,{\rm
GeV}^2$.
We note that in contrast to the lepton spectra, there are $1/m_b$
corrections both to the  physical   hadronic invariant mass spectrum and to
the physical hadronic energy spectrum, although these corrections are
absent for the $\hat s_0$ and $\hat E_0$ spectra \cite{BU}.

While the complete shape of the $\hat E_0$ spectrum may be calculated
(away from the parton model endpoint $\hat E_0={1\over2}(1+\hat m_q^2)$)
with the standard OPE analysis, only suitably averaged features, such as
moments, of the
$\hat s_0$ spectrum may be computed reliably.  The difference arises
because each point of   the $\hat E_0$ spectrum receives contributions from
states of different invariant masses, making the   process inclusive,
whereas
by definition each point of the $\hat s_0$ spectrum only receives
contributions
from states of a single invariant mass.   This may be seen explicitly by
carrying out the   usual OPE analysis for inclusive decays in the variables
$\hat s_0$ and $\hat E_0$, instead of the usual leptonic variables
$v\cdot\hat q$ and $\hat q^2$.

The inclusive $B$ meson decay rate is given by
\begin{equation}\label{contour}
   \Gamma(B\to X_q\ell\nu)\sim\int{\rm d}\hat s_0\, {\rm d}\hat E_0\,
   \sqrt{\hat E_0^2 - \hat s_0}
   L_{\mu\nu}(\hat s_0,\hat E_0)W^{\mu\nu}(\hat s_0,\hat E_0)\,,
\end{equation}
where $L_{\mu\nu}$ is the spin summed lepton tensor
$L_{\mu\nu} \propto ( q_\mu q_\nu - g_{\mu\nu} q^2 )$. Using the optical
theorem, the nonperturbative hadronic tensor
$W^{\mu\nu}$ is related to the imaginary part of the forward scattering
amplitude \cite{SV,CGG},
\begin{eqnarray}
   W^{\mu\nu} &=& \sum_X \langle B|\,J_h^{\mu\dagger}\,|X\rangle
   \langle X|\,J_h^\nu\,|B\rangle (2\pi)^4\delta^4(P_B - P_X-q)\nonumber\\
   &=& -2\,{\rm Im}\,\langle B|\,i\int{\rm d}x\,e^{-iq\cdot x}
   \,T \left[  J_h^{\mu\dagger}(x) J_h^\nu(0)  \right]\,|B\rangle\\
   &\equiv& 2\,{\rm Im}\,T^{\mu\nu}\,\nonumber
\end{eqnarray}
where $J^\mu_h = \overline{q} \gamma^\mu {1\over 2}(1-\gamma_5)  b$.  The
time-ordered product $T^{\mu\nu}$ may be written via an  operator product
expansion as a power series in $\alpha_s(m_b)$ and $1/m_b$.

\begin{figure}
\epsfxsize=15cm
\hfil\epsfbox{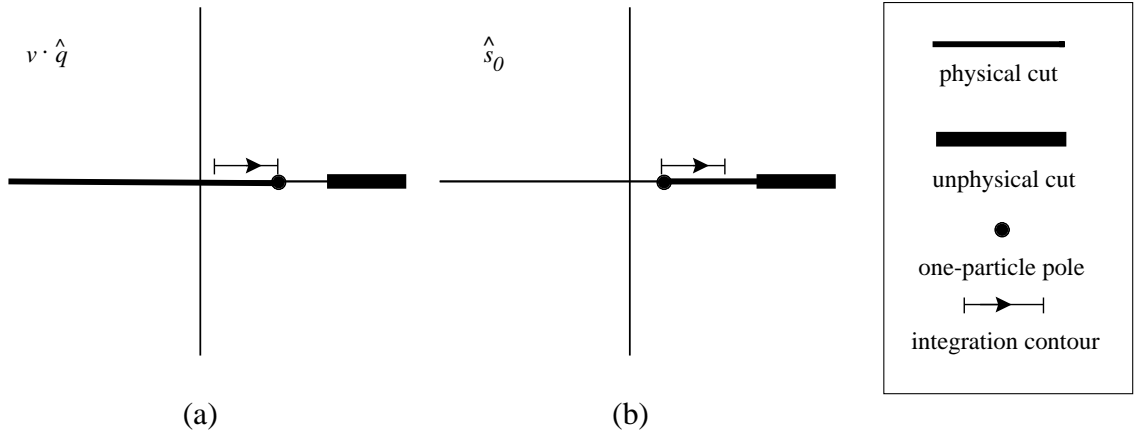}\hfill
\caption{The analytic structure of $T^{\mu\nu}$, (a) in the $v\cdot\hat q$
plane, with $\hat q^2$ fixed, and (b) in the $\hat s_0$ plane, with $\hat
E_0$
fixed.  Both the physical and unphysical cuts
are shown, as well as the position of the one-particle pole.}
\label{analytic}
\end{figure}

In the $v\cdot \hat q$ plane, for fixed $\hat q^2$, the  correlator
$T^{\mu\nu}$ has the analytic structure shown in Fig.~\ref{analytic}a, as
discussed in
Ref.~\cite{CGG}.
There are cuts along the real axis, a physical one
(corresponding to $B$ decays) for
$v\cdot\hat q\le{1\over2}[1+\hat q^2-\hat m_q^2]$, and an  unphysical   one
(corresponding to scattering processes) for
$v\cdot\hat q\ge{1\over2}[(2+\hat m_q)^2-\hat q^2-1]$.  The   one-particle
pole lies at the right hand end of the physical cut.  After an integral
over the charged  lepton   energy, the decay rate is computed by performing
an integration over the top of the physical cut,  for
$\sqrt{\hat q^2}\le v\cdot\hat q\le{1\over2}(1+\hat q^2-\hat m_q^2)$,
followed by an integration
over $0\le\hat   q^2\le(1-\hat m_q)^2$.  Note that in the limit $\hat
m_q\to0$ and
$\hat q^2\to1$, the physical and unphysical cuts pinch the region  of
integration.  In this corner of the parameter space, the operator product
expansion breaks down.  Attempts to resum the OPE to all orders in
this region have thus far proven inconclusive~\cite{MW,Neubert,FJMW}.

Mapping from the $v\cdot\hat q$ plane to the $\hat s_0$ plane at   fixed
$\hat E_0$, one finds two cuts on the positive
$\hat s_0$ axis, as shown in Fig.~\ref{analytic}b.  The physical cut, which
terminates in the one-particle pole, extends over
$\hat s_0\ge\hat m_q^2$.  The unphysical cut lies away from the   pole, at
$\hat s_0\ge\hat m_q^2+4\hat m_q+4\hat E_0$.  The region of integration in
$\hat  s_0$   is given by ${\rm max}(2\hat E_0-1,\hat m_q)
\le\hat s_0\le\hat E_0^2$, to be followed by an integration  in
$\hat E_0$, over $\hat m_q\le\hat E_0\le1$.

In these variables, the region of integration only touches the  unphysical
cut in the limit $\hat m_q\to0$ and $\hat E_0\to0$, which from
Eq.~(\ref{relations}) is   equivalent to the condition for the cuts to
pinch in the $v\cdot\hat q$ plane.  In this singular   region, as before,
the operator product expansion breaks down.  We also note that the
integration   region covers the one-particle pole at $\hat s_0=\hat m_q^2$
only if $\hat  E_0\le{1\over2}(1+\hat   m_q^2)$.  Indeed, this corresponds
to the maximum energy the final quark can take away  in   the decay
process.  The cut for $\hat s_0>\hat m_q^2$ is populated only by
multiparticle  final   states generated by the radiation of gluons.  In
perturbation theory, then, the  differential   spectrum
${\rm d}\Gamma/{\rm d}\hat E_0$ for $\hat E_0>{1\over2}(1+\hat m_q^2)$  is
of order $\alpha_s$.

As is the case for $\tau$ decays, the contour of integration in
Eq.~(\ref{contour}) may be deformed away from the physical region,
except at the point the contour
crosses the physical cut.  However,
we note that in contrast to $\tau$ decays, the integrand in
Eq.~(\ref{contour})
does not have a double zero where the deformed contour approaches the
physical region.  It is possible,
therefore, that deviations from quark/hadron duality in the Minkowski
regime
may be more pronounced in semileptonic heavy hadron decay than in $\tau$
decay.

\section{Spectral Moments}

In this section we compute the spectral moments at the parton level.  We
will
treat both the leading power corrections, proportional to $\lambda_1$ and
$\lambda_2$, and the leading perturbative contributions, proportional to
$\alpha_s(m_b)$.  We take the two types of corrections in turn.

\subsection{Power Corrections}

For the computation of the power corrections, it is convenient to
decompose the
time ordered product $T^{\mu\nu}$ into the
form factors
\begin{equation}
   T^{\mu\nu}(\hat s_0,\hat E_0)=-g^{\mu\nu}T_1(\hat s_0,\hat E_0)
   +v^\mu v^\nu T_2(\hat s_0,\hat E_0)+\dots\,,
\end{equation}
where the omitted form factors vanish for massless leptons in the final
state.
In terms of $T_1$ and $T_2$ the differential spectrum is given by
\begin{equation}\label{diffspect}
  {1\over\Gamma_0}{{\rm d}\Gamma\over{\rm d}\hat s_0{\rm d}\hat E_0}=
  -{32\over\pi}{\rm Im}\sqrt
  {\hat E_0^2-\hat s_0}\left[3(1-2\hat E_0+\hat s_0) T_1(\hat s_0,\hat E_0)
  +(\hat E_0^2-\hat s_0) T_2(\hat s_0,\hat E_0)\right]\,,
\end{equation}
where
\begin{equation}
  \Gamma_0={G_F^2 m_b^5|V_{bq}|^2\over 192\pi^3}
\end{equation}
is proportional to the total decay rate.

The leading $1/m_b$ corrections to the hadronic quantities $T_1$  and
$T_2$ were calculated in Refs.~\cite{Bigietc,MW}. In terms of $\hat s_0$
and $\hat E_0$, they are given by
\begin{eqnarray}\label{t1def}
  T_1(\hat s_0,\hat E_0)=&&
  {1\over\hat s_0-{\hat m}_q^2+i\epsilon}\left[{\hat E_0\over 2}-
  {\lambda_1\over 12 m_b^2}-{\lambda_2\over 4m_b^2}\right]\nonumber  \\
  &&\mbox{}+{1\over\left(\hat s_0-{\hat m}_q^2+i\epsilon\right)^2}
  \left[{\lambda_1\over6m_b^2}\left(
  5\hat E_0^2-3\hat E_0-2\hat s_0\right)+
  {\lambda_2\over2
  m_b^2}\left(5\hat E_0^2+\hat E_0-2\hat s_0\right)\right] \\
  &&\mbox{}+{1\over\left(\hat s_0-{\hat m}_q^2+i\epsilon\right)^3}
  \left[{2\lambda_1\over 3m_b^2}\hat E_0(\hat s_0-
  \hat E_0^2)\right]\,,\nonumber\\
  T_2(\hat s_0,\hat E_0)=&&{1\over\hat s_0-{\hat m}_q^2+i\epsilon}
  \left[1-{5\lambda_1\over 6m_b^2}
  -{5\lambda_2\over 2m_b^2}\right] \nonumber \\
  &&\mbox{}+{1\over\left(\hat s_0-{\hat m}_q^2+i\epsilon\right)^2}
  \left[{7\lambda_1\over 3
  m_b^2}(\hat E_0-1)+{\lambda_2\over m_b^2}(5\hat E_0-3)\right] \\
  &&\mbox{}+{1\over\left(\hat s_0-{\hat m}_q^2+i\epsilon\right)^3}
  \left[{4\lambda_1\over
  3m_b^2}(\hat s_0-\hat E_0^2)\right]\,.
  \nonumber
\end{eqnarray}
Integrating this expression with respect to $\hat E_0$, we find the leading
power correction to the invariant mass spectrum.  Of course, since there is
only a single quark in the final state, this expression is a singular
function
with support only at $\hat s_0-\hat m_q^2$.  Only its moments, which we
present below, are meaningful.  The corrections to the hadronic
energy spectrum, obtained by integrating first with respect to $\hat s_0$,
are
more interesting, and are presented in Appendix~A.

Because the expansions of $T_1$ and $T_2$ in terms of
$1/m_b$ contain pole factors $1/(\hat s_0-\hat m_q^2)^n$, it is simplest
to compute the moments of $(\hat s_0-\hat m_q^2)$ rather than those of
$\hat s_0$.  The requisite calculations are  straightforward but tedious,
and we present only the final results.  It is convenient to scale the
various contributions to
$\Gamma_0$ rather than to the full width $\Gamma$; the quantities which we
will
present below are then of the form
\begin{equation}
{\cal M}^{(n,m)} =
   {1\over\Gamma_0}\int (\hat s_0-\hat m_q^2)^n \hat E_0^m\,
   {{\rm d}\Gamma\over{\rm d}\hat s_0{\rm d}\hat E_0}
   \,{\rm d}\hat s_0{\rm d}\hat E_0\,,
\end{equation}
for integers $n$ and $m$.  They are related to the parton level moments by
a
scaling to the corrected decay rate,
\begin{equation}
   \langle\hat E_0^m(\hat s_0-\hat m_q^2)^n\rangle =
{\Gamma_0\over\Gamma}\,
   {\cal M}^{(n,m)}\,.
\end{equation}

\subsection{Perturbative Corrections}

\begin{figure}
\epsfxsize=10cm
\hfil\epsfbox{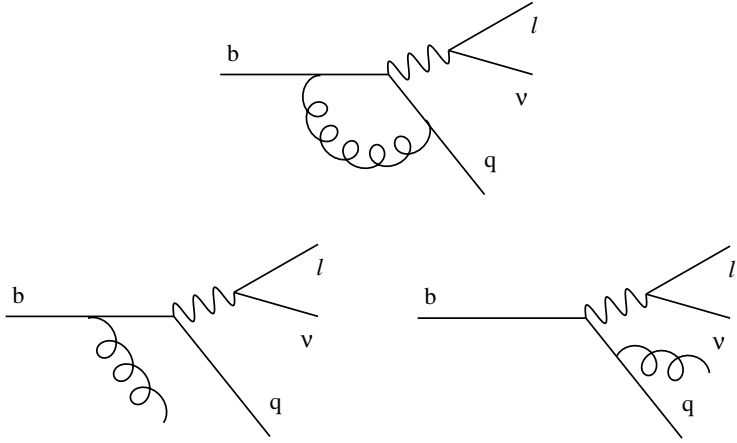}\hfill
\caption{The Feynman diagrams which contribute to the moments at
order $\alpha_s$.  There are also wave function corrections, which we do
not
show.}
\label{graphs}
\end{figure}

The perturbative corrections to $T_1$ and $T_2$ are most
conveniently calculated directly from the graphs in Fig.~\ref{graphs}.
The radiative contributions to $\langle(\hat s_0-\hat m_q^2)^n\rangle$ come
only from bremsstrahlung graphs and are straightforward to compute for
arbitrary $\hat m_q$.  We find
\begin{eqnarray}\label{sspectrad}
  {1\over\Gamma_0}{{\rm d}\Gamma\over {\rm d}\hat
  s_0}&=&{\alpha_s\over \pi}{1\over\hat s_0-\hat m_q^2}
  \Bigg[{(\hat s_0-1)\over 27 \hat s_0^2}\left(-9 \hat m_q^4-6\hat m_q^6+
  \hat s_0(18\hat m_q^2+81\hat m_q^4+48\hat m_q^6)\right. \\ \nonumber
  &&\qquad\qquad\mbox{}+\hat s_0^2(93-316\hat m_q^2+243 \hat m_q^4+102\hat
  m_q^6)+
  \hat s_0^3(-41-478\hat m_q^2+9\hat m_q^4) \\ \nonumber
  &&\qquad\qquad\mbox{}+\left.\hat s_0^4(-95-64\hat m_q^2)+55\hat
s_0^5\right)
  \\
  \nonumber
  &&\qquad\mbox{}+{4\over 9}\ln\hat s_0\left(-3+5\hat m_q^2-18\hat m_q^4
  -9\hat m_q^6+\hat s_0(-5+45\hat m_q^2-9\hat m_q^4-3\hat m_q^6)\right. \\
  & &\qquad\qquad\mbox{} +9\left.\hat s_0^2(1+2\hat m_q^2)+ 2\hat m_q^2
\hat
  s_0^3-2\hat s_0^4\right)\Bigg]
  \,,\nonumber
\end{eqnarray}
from which it is easy to extract the moments
$\langle(\hat s_0-\hat m_q^2)^n\rangle$.  Similarly, weighting with extra
factors of $\hat E_0$ yields the radiative correction to the moments ${\cal
M}^{(n,m)}$, for $n\ge1$ and any $m$.

The one-loop radiative corrections to the hadronic energy spectrum, and
hence
to the moments ${\cal M}^{(0,m)}$, are considerably more difficult to
compute.
This is because they receive contributions from both virtual graphs and
bremsstrahlung graphs, only the sum of which is infrared finite.  The
complete
calculation of the radiative corrections to the differential energy
spectrum
${\rm d}\Gamma/{\rm d}\hat E_0$ was computed by Czarnecki, Je\.zabek and
K\"uhn~\cite{CJK}.

We present the leading perturbative corrections to $\langle\hat E_0\rangle$
and $\langle\hat E_0^2\rangle$ in Appendix~A.  In the limit
$\hat m_q\to 0$ they take the simple form
\begin{eqnarray}\label{e0pert}
   {\cal M}^{(0,1)}_{\rm pert.}(\hat m_q=0)
   &=&\bigg[{1381\over900}
   -{7\over30}\pi^2\bigg]{\alpha_s\over\pi}
   =-0.768\,{\alpha_s\over\pi}\,,\nonumber\\
   {\cal M}^{(0,2)}_{\rm pert.}(\hat m_q=0)
   &=&\bigg[{2257\over3600}
   -{4\over45}\pi^2\bigg]{\alpha_s\over\pi}
   =-0.250\,{\alpha_s\over\pi}\,.
\end{eqnarray}

\subsection{Corrections to the Moments}

We now combine the results of the previous subsections to present the full
expressions for the parton-level moments, including the leading
perturbative
and power corrections.

The first two moments of the hadronic invariant mass spectrum are
given by
\begin{eqnarray}\label{s0np}
  {\cal M}^{(1,0)} &=&
   {\alpha_s\over\pi}\left[{91\over450}+{71 \over 18}\hat m_q^2-
   {158 \over 27}\hat m_q^4+{34\over 9}\hat m_q^6+{1\over 18}\hat m_q^8-
   {2873\over 1350}\hat m_q^{10}\right.  \nonumber \\
   &&\qquad\mbox{}+\left.\left(4\hat m_q^2+
   {40\over 9}\hat m_q^4-{8\over 3}\hat m_q^6+
   {20\over 3}\hat m_q^8+{56\over 45}\hat m_q^{10}\right)\ln\hat
m_q\right] \\
   &&\mbox{}+{\lambda_1\over2m_b^2}\bigg[{13\over10}-{27\over2}\hat m_q^2-
   16\hat m_q^4
   +32\hat m_q^6-{9\over2}\hat m_q^8+{7\over10}\hat m_q^{10}-60\hat
   m_q^4\ln\hat m_q-12\hat m_q^6\ln\hat
   m_q\bigg] \nonumber \\
   &&\mbox{}+{\lambda_2\over2m_b^2}\bigg[{3\over2}-{9\over2}\hat m_q^2+
   8\hat m_q^4
   -24\hat m_q^6+{45\over2}\hat m_q^8-{7\over2}\hat m_q^{10}+12\hat
   m_q^4\ln\hat m_q-36\hat m_q^6
   \ln\hat m_q\bigg]\,,\nonumber\\
   {\cal M}^{(2,0)} &=&
   {\alpha_s\over\pi}\left[{5\over324}-{137 \over 450}\hat m_q^2-
   {101 \over 36}\hat m_q^4+{86\over81}\hat m_q^6-{29\over
   36}\hat m_q^8+{37\over
   18}\hat m_q^{10}+{6341\over 8100}\hat m_q^{12}\right.  \\
   \nonumber
   &&\qquad-\left.\left({10\over 3}\hat m_q^4+{152\over 27}\hat m_q^6+
   {14\over 3}\hat m_q^8+4\hat m_q^{10}+{56\over
   135}\hat m_q^{12}\right)\ln\hat m_q\right]  \\
   &&\mbox{}+{\lambda_1\over2m_b^2}\bigg[-{16\over45}+{16\over5}
   \hat m_q^2-16\hat m_q^4
   +16\hat m_q^8-{16\over5}\hat m_q^{10}+{16\over45}\hat m_q^{12}
   -{128\over3}\hat m_q^6\ln\hat m_q\bigg]\,. \nonumber
\end{eqnarray}
The first mixed moment is
\begin{eqnarray}\label{e0s0np}
   {\cal M}^{(1,1)} &=&
   {\alpha_s\over\pi}\left[{9\over100}+{209 \over 180}\hat m_q^2-
   {149 \over 108}\hat m_q^4+{4\over 3}\hat m_q^6-{49\over 36}\hat m_q^8
   +{1457\over 2700}\hat m_q^{10}-{23\over 60}\hat m_q^{12}\right. \\
   \nonumber
   &&\qquad\mbox{}+\left.\left({7\over 5}\hat m_q^2+{10\over
   9}\hat m_q^4+{4\over 3}\hat m_q^6+
   {2\over 3}\hat m_q^8+{23\over 45}\hat m_q^{10}
   +{4\over 15}\hat m_q^{12}\right)
   \ln\hat m_q\right]\nonumber \\
   &&\mbox{}+{\lambda_1\over2m_b^2}\bigg[{23\over90}-{1\over2}\hat
   m_q^2+12\hat m_q^4
   -16\hat m_q^6+{13\over2}\hat m_q^8-{27\over10}\hat
   m_q^{10}+{4\over9}\hat m_q^{12}\nonumber\\
   &&\qquad\qquad\qquad+12\hat m_q^4\ln\hat m_q+{20\over3}\hat m_q^6\ln\hat
   m_q\bigg]\nonumber \\
   &&\mbox{}+{\lambda_2\over2m_b^2}\bigg[{13\over30}+{3\over2}\hat
   m_q^2-4\hat m_q^4
   -{3\over2}\hat m_q^8+{49\over10}\hat m_q^{10}-{4\over3}\hat
   m_q^{12}+12\hat m_q^4\ln\hat m_q-20\hat
   m_q^6\ln\hat m_q\bigg]
   \,.\nonumber
\end{eqnarray}
The first two moments of the hadron energy spectrum are given by
\begin{eqnarray}\label{e0np}
{\cal M}^{(0,1)}&=&
   \bigg[{7\over20}-{5\over4}\hat m_q^2+8\hat m_q^4-8\hat m_q^6
   +{5\over4}\hat m_q^8-{7\over20}\hat m_q^{10}
   +6\hat m_q^4\ln\hat m_q+6\hat m_q^6\ln\hat m_q\bigg]\nonumber\\
   &&\mbox{}+A_1(\hat m_q)\,{\alpha_s\over\pi}\\
   &&\mbox{}+{\lambda_1\over2m_b^2}\,\bigg[1-8\hat m_q^2+8\hat m_q^6
   -\hat m_q^8-24\hat m_q^4\ln\hat m_q\bigg]\nonumber\\
   &&+{\lambda_2\over2m_b^2}\,\bigg[7\hat m_q^2-20\hat m_q^4
   +20\hat m_q^8-7\hat m_q^{10}+24\hat m_q^2\ln\hat m_q
   -48\hat m_q^4\ln\hat m_q\bigg]\,,\nonumber\\
   {\cal M}^{(0,2)}&=&
   \bigg[{2\over15}-{1\over5}\hat m_q^2-2\hat m_q^4
   +2\hat m_q^8+{1\over5}\hat m_q^{10}-{2\over15}\hat m_q^{12}
   -8\hat m_q^6\ln\hat m_q\bigg]\nonumber\\
   &&\mbox{}+A_2(\hat m_q)\,{\alpha_s\over\pi}\\
   &&\mbox{}+{\lambda_1 \over 2m_b^2} \, \bigg[{43\over 90}
   -{3\over2}\hat m_q^2+14\hat m_q^4-16\hat m_q^6
   +{9\over2}\hat m_q^8-{17\over10}\hat m_q^{10}
   +{2\over9}\hat m_q^{12} \nonumber\\
   &&\qquad\qquad\qquad\mbox{}+12\hat m_q^4\ln\hat m_q
   +{28\over3}\hat m_q^6\ln\hat m_q\bigg]\nonumber \\
   &&\mbox{}+{\lambda_2\over 2m_b^2}\,\bigg[{13\over30}
   -{21\over 2}\hat m_q^2+50\hat m_q^4-52\hat m_q^6
   +{21\over 2}\hat m_q^8+{49\over 10}\hat m_q^{10}
   -{10\over 3}\hat m_q^{12}\nonumber\\
   &&\qquad\qquad\qquad\mbox{}+12\hat m_q^4\ln\hat m_q
   +28\hat m_q^6\ln\hat m_q\bigg]\,,\nonumber
\end{eqnarray}
where the functions $A_m(\hat m_q)$ are presented in Appendix A.

Finally, we obtain the leading corrections to the total decay rate
by taking the $n=m=0$ moment.  Of course, this result is not new; we
present it
for completeness and because we will need it to normalize the moments.  We
find
\begin{equation}\label{totalwidth}
  \Gamma(B\to X_q e\bar\nu)=\Gamma_0\left[
  f_0(\hat m_q)+{1\over 2 m_b^2}f_1(\hat
  m_q,\lambda_1,\lambda_2)+A_0(\hat m_q)\,{\alpha_s\over\pi}\right]\,,
\end{equation}
where
\begin{eqnarray}
  &&f_0(\hat m_q)=1-8\hat m_q^2+8\hat m_q^6-\hat m_q^8-24\hat m_q^4
  \ln\hat m_q\,,\nonumber\\
  &&f_1(\hat m_q,\lambda_1,\lambda_2)=\lambda_1\left(1-8\hat m_q^2+8\hat
  m_q^6
  -\hat m_q^8-24\hat m_q^4\ln\hat m_q\right) \\
  &&\qquad\qquad\qquad\qquad\mbox{}+\lambda_2\left(-9+24\hat m_q^2-72\hat
  m_q^4+72\hat m_q^6-15\hat m_q^8-72\hat m_q^4\ln\hat
m_q\right)\,.\nonumber
\end{eqnarray}
The power correction $f_1(\hat m_q,\lambda_1,\lambda_2)$ was first
obtained in
Refs.~\cite{Bigietc,MW}, and the perturbative correction $A_0(\hat m_q)$,
which
we present in Appendix A, was first found in Ref.~\cite{Nir}.  It takes a
simple form when $\hat m_q\to0$, for which
\begin{equation}
   A_0(\hat m_q=0) = {25\over6}-{2\over3}\pi^2\,.
\end{equation}

\section{Application to $B$ Meson Decays}

The relations (\ref{physparton}) allow  moments of  the physical
parameters $E_H$ and $s_H$ to be expressed in terms of the  parton-level
moments.   For the first two moments of $s_H$ we find
\begin{eqnarray}\label{sHmoments}
   \langle s_H\rangle&=&m_q^2+\bar\Lambda^2+(m_B^2-2\bar\Lambda
   m_B+\bar\Lambda^2
  +\lambda_1+3\lambda_2)\,\langle\hat s_0-\hat
  m_q^2\rangle\nonumber\\
  &&\quad\mbox{}+(2\bar\Lambda m_B-2\bar\Lambda^2-\lambda_1-3\lambda_2)
  \langle\hat E_0\rangle\,,\nonumber\\
  \langle s_H^2\rangle&=&m_q^4+2\bar\Lambda^2m_q^2+
  2m_B^2 (m_q^2+\bar\Lambda^2)
  \langle\hat s_0-\hat m_q^2\rangle\nonumber\\
  &&\quad\mbox{}+2m_q^2(2\bar\Lambda
  m_B-2\bar\Lambda^2-\lambda_1-3\lambda_2)\langle\hat E_0\rangle
  \nonumber\\
  &&\quad\mbox{}+ (m_B^4-4\bar\Lambda m_B^3+6\bar\Lambda^2 m_B^2
  +2\lambda_1 m_B^2+6\lambda_2 m_B^2)\langle (\hat s_0-\hat
  m_q^2)^2\rangle\\
  &&\quad\mbox{}+ 4\bar\Lambda^2 m_B^2 \langle\hat E_0^2\rangle+
  4\bar\Lambda m_B^3\langle\hat E_0(\hat s_0-\hat m_q^2)\rangle\,,\nonumber
\end{eqnarray}
where all expressions are valid to relative order $\Lambda_{\rm
QCD}^2/m_b^2$
and to all orders in $m_q/m_b$.  It is
straightforward to extend the analysis to higher moments $\langle
s_H^n\rangle$.  Similarly, the leading moments of $E_H$ are given by
\begin{eqnarray}\label{EHmoments}
  \langle E_H\rangle &=& \bar\Lambda-{\lambda_1+3\lambda_2\over2m_B}
  +\left(m_B-\bar\Lambda+{\lambda_1+3\lambda_2\over2m_B}\right)\langle
  \hat E_0\rangle\,,\nonumber\\
  \langle E_H^2\rangle &=& \bar\Lambda^2 + (2\bar\Lambda m_B -
2\bar\Lambda^2
  -\lambda_1-3\lambda_2)\langle \hat E_0\rangle\\
  &&\quad +(m_B^2-2\bar\Lambda m_B+\bar\Lambda^2+\lambda_1+3\lambda_2)
  \langle\hat E_0^2\rangle\,.\nonumber
\end{eqnarray}
On the right hand side of these expressions appear the parton level moments
$\langle \hat E_0^m\hat s_0^n\rangle$, which are obtained from the
quantities
${\cal M}^{(n,m)}$ by multiplying by the scale factor $\Gamma_0/\Gamma$.

\subsection{Decay to an up quark}

For $\hat m_q=0$, such as in the quark decay $b\to u\ell\bar\nu$, we find
the
simple expressions
\begin{eqnarray}\label{masslessone}
   \langle \hat
s_0\rangle&=&{91\over450}{\alpha_s\over\pi}+{13\lambda_1\over
   20 m_B^2}+{3\lambda_2\over 4 m_B^2}\,,\nonumber \\
   \langle \hat
s_0^2\rangle&=&{5\over324}{\alpha_s\over\pi}-{16\lambda_1\over
   90 m_B^2}\,, \\
   \langle \hat E_0 \rangle&=&{7\over
   20}\left(1+{137\over 630}
   {\alpha_s\over\pi}+{13\lambda_1\over
   14 m_B^2}+{9\lambda_2\over 2 m_B^2}\right)\,,  \nonumber \\
   \langle \hat E_0^2 \rangle&=&{2\over
   15}\left(1+{257\over 480}
   {\alpha_s\over\pi}+{31\lambda_1\over
   24 m_B^2}+{49\lambda_2\over 8 m_B^2}\right)\,,
   \nonumber \\
   \langle \hat E_0 \hat
   s_0\rangle&=&{9\over100}{\alpha_s\over\pi}+{23\lambda_1
   \over180 m_B^2}+{13\lambda_2\over 60 m_B^2}\,, \nonumber
\end{eqnarray}
accurate up to corrections of order
$\alpha_s/m_B^2$ and $1/m_B^3$.
These then yield the physical moments
\begin{eqnarray}\label{masslesstwo}
    \langle s_H \rangle&=&m_B^2\left[{91\over 450}
    {\alpha_s\over\pi}
    +{7\bar\Lambda\over 10 m_B}\left(1-{227\over 630}
    {\alpha_s\over\pi}\right)
    +{3\over10 m_B^2}\left(\bar\Lambda^2+\lambda_1-\lambda_2
    \right)\right]\,,  \nonumber \\
    \langle s_H^2 \rangle&=&m_B^4\left[{5\over 324}{\alpha_s\over\pi}+
  {604\bar\Lambda\over 2025 m_B}{\alpha_s\over\pi}+
  {8\over 15m_B^2}\left(\bar\Lambda^2-{\lambda_1\over 3}\right)\right]\,,\\
  \langle E_H\rangle &=&
{7\over20}m_B\left[1+{137\over630}{\alpha_s\over\pi}
  +{13\bar\Lambda\over7m_B}\left(1-{137\over1170}{\alpha_s\over\pi}\right)
  +{12\lambda_2\over7m_B^2}\right]\,,
    \nonumber\\
    \langle E_H^2\rangle &=& {2\over15}m_B^2\left[1+{257\over480}
    {\alpha_s\over\pi}+{13\bar\Lambda\over4m_B}\left(1+{17\over780}
    {\alpha_s\over\pi}\right)+{13\over4m_B^2}\left(\bar\Lambda^2
    -{4\over39}\lambda_1+{5\over 13}\lambda_2\right)\right]\,.\nonumber
\end{eqnarray}

\subsection{Decay to a charm quark}

For $b\rightarrow c$ decays, we make use of the fact that the charm quark
is also heavy
to write $m_c/m_b$ as a power series in $1/m_B, 1/m_D$.  Let us define the
spin-averaged meson masses,
\begin{eqnarray}\label{massratio}
  \bar m_D\equiv{m_D+3 m_{D^*}\over 4}&=&m_c+\bar\Lambda-{\lambda_1\over
  2m_D}+\ldots\simeq 1975\,{\rm MeV}\, \\
  \bar m_B\equiv{m_B+3 m_{B^*}\over 4}&=&m_b+\bar\Lambda-{\lambda_1\over
  2m_B}+\ldots\simeq 5313\,{\rm MeV}\,,\nonumber
\end{eqnarray}
which gives
\begin{eqnarray}
  {m_c\over m_b}&=&{\bar m_D\over \bar m_B}-{\bar\Lambda\over m_B}\left(
  1-{\bar m_D\over \bar m_B}\right)-{\bar\Lambda^2\over m_B^2}\left(
  1-{\bar m_D\over \bar m_B}\right)+{\lambda_1\over 2 m_B m_D}\left(
  1-{\bar m_D^2\over \bar m_B^2}\right) \\
  &=&0.372-0.628{\bar\Lambda\over m_B}-0.628{\bar\Lambda^2\over
  m_B^2}+1.16{\lambda_1\over m_B^2}\,, \nonumber
\end{eqnarray}
accurate up to corrections of order $1/m_B m_D^2$.
This substitution introduces additional ${\cal O}(1/m_B, 1/m_B m_D)$
corrections to the parton level
moments.\footnote{In the rest of this section, we will treat $m_B/m_D$ as
${\cal O}(1)$.  Thus, by ${\cal O}(1/m_B)$ we denote corrections both of
order
$1/m_B$ and $1/m_D$.}  We find
\begin{eqnarray}\label{massiveone}
   \langle\hat s_0-\hat m_q^2\rangle=&&0.051{\alpha_s\over\pi}+
   0.16{\alpha_s\over\pi}{\bar\Lambda\over m_B}+0.51
   {\lambda_1 \over m_B^2}+1.14 {\lambda_2\over m_B^2}\,,\nonumber\\
   \langle(\hat s_0-\hat m_q^2)^2\rangle=&&0.0053{\alpha_s\over\pi}+
   0.017{\alpha_s\over\pi}{\bar\Lambda\over m_B}
   -0.14{\lambda_1 \over m_B^2}\,,\\
   \langle\hat E_0\rangle=0.489&&\left[1+0.043{\alpha_s\over\pi}-
   0.78{\bar\Lambda\over m_B}\left(1-0.12{\alpha_s\over \pi}\right)
   -0.44{\bar \Lambda^2\over m_B^2}+1.96{\lambda_1\over m_B^2}+2.53
   {\lambda_2\over m_B^2}\right]\nonumber\\
   \langle\hat E_0^2\rangle=0.242&&\left[1+0.099{\alpha_s\over\pi}-
   1.50{\bar\Lambda\over m_B}\left(1-0.12{\alpha_s\over \pi}\right)
   -0.19{\bar \Lambda^2\over m_B^2}+3.64{\lambda_1\over m_B^2}+4.69
   {\lambda_2\over m_B^2}\right]\, \nonumber \\
   \langle\hat E_0(\hat s_0-\hat
    m_q^2)\rangle=&&0.030{\alpha_s\over\pi}+
   0.077{\alpha_s\over\pi}{\bar\Lambda\over m_B}+0.18
   {\lambda_1 \over m_B^2}+0.53 {\lambda_2\over m_B^2}\,,\nonumber
\end{eqnarray}
and for the total rate,
\begin{equation}\label{gammaexpr}
   {\Gamma\over\Gamma_0}=0.369\left[1-1.54{\alpha_s\over\pi}+
    3.35 {\bar\Lambda\over m_B}\left(1-1.86{\alpha_s\over\pi}\right)+
    5.81{\bar\Lambda^2\over m_B^2}-5.69{\lambda_1\over m_B^2}-
    7.47{\lambda_2\over m_B^2}\right]\,.
\end{equation}
The physical moments are then
\begin{eqnarray}\label{massivetwo}
   \langle s_H-\bar m_D^2\rangle=&&m_B^2\left[0.051{\alpha_s\over\pi}
   +0.23{\bar\Lambda\over m_B}\left(1+0.43{\alpha_s\over\pi}
   \right)+0.26{1\over  m_B^2}\left(\bar\Lambda^2
   +3.9\lambda_1-1.2\lambda_2\right)\right],\nonumber \\
  \langle (s_H^2-\bar m_D^2)^2\rangle=&&m_B^4\left[0.0053{\alpha_s\over\pi}
   +0.067{\bar\Lambda\over m_B}{\alpha_s\over\pi}
   +0.065{1\over m_B^2}\left(\bar\Lambda^2-2.1\lambda_1\right)\right] \\
   \langle E_H\rangle=0.489m_B&&\left[1+0.043{\alpha_s\over\pi}
   +0.27{\bar\Lambda\over m_B}\left(1+0.19
   {\alpha_s\over\pi}\right)+0.33{1\over m_B^2}\left(\bar\Lambda^2+
   4.3\lambda_1+2.9\lambda_2\right)\right]\,,\nonumber\\
   \langle E_H^2\rangle=0.242m_B^2&&\left[1+0.099{\alpha_s\over\pi}
   +0.55{\bar\Lambda\over m_B}\left(1+0.28
   {\alpha_s\over\pi}\right)+0.75{1\over m_B^2}\left(\bar\Lambda^2
   +3.5\lambda_1+2.2\lambda_2\right)\right],\nonumber
\end{eqnarray}
where the corrections to these expressions are of order $\alpha_s/m_B^2$
and
$1/m_B^3$.  Note that in these expansions, there is no hidden dependence
on the
quark masses; here the coefficients are functions {\it only\/} of physical
quantities.

We can also expand our results about the small velocity (SV)
limit \cite{SVLimit}, $\Lambda_{\rm QCD}\ll m_b-m_c\ll m_c<m_b$.  In  this
limit, only the $D$ and $D^*$ states are produced.  Expanding
in powers of $1-\hat m_c$ and $\Lambda_{\rm QCD}/(m_b-m_c)$, we find
\begin{equation}
    \langle s_H-\bar m_D^2\rangle=m_B^2\,(1-\hat m_c)^3\,\left[
     {4\over 21}{\alpha_s\over\pi}+{4\bar\Lambda\over
     m_b-m_c}+{\lambda_2-2\lambda_1\over
     (m_b-m_c)^2}+\dots \right]+{\cal O}(1-\hat m_c)^4.
\end{equation}
Note that in the SV limit, as expected, the average invariant mass of
the final hadron is $\bar m_D$, and therefore the $D$ and $D^*$ are
produced in the ratio 1:3.  Furthermore, corrections to the average
invariant
mass due to production of excited states are suppressed by $(1-\hat
m_c)^3$.

Finally, all of these results may be applied to the inclusive decays of the
$\Lambda_b$, with the obvious replacements $\bar\Lambda\rightarrow
\bar\Lambda_{\Lambda}, \lambda_1\rightarrow \lambda_{1\Lambda},
\ \lambda_2\rightarrow 0$, where
\begin{equation}
   m_{\Lambda_b}=m_b+\bar\Lambda_\Lambda-{\lambda_{1\Lambda}\over 2m_b}
   +\dots\, .
\end{equation}
We also note that, in order to avoid introducing factors of $\bar\Lambda$
and
$\lambda_1$ from the meson sector into the expansion, in Eq.
(\ref{massratio})
the spin-averaged meson masses should be replaced by baryon masses.  Since
the uncertainty in $m_{\Lambda_b}$ is $\pm 50$ MeV \cite{PDG}, this
introduces
large uncertainties into the moments of $\Lambda_b$ spectra, when written
in
terms of physical masses.

\section{A Lower Bound on $\bar\Lambda$}

Although the invariant mass spectrum for $B\rightarrow X_c  e\bar\nu$
has not been measured, we
may use the recent OPAL measurement \cite{opal}  of the branching ratio to
the narrow
$P$ wave charmed mesons, the $D_1(2420)$ and $D_2^*(2460)$, to place a
lower limit on
$\bar\Lambda$.  In Ref.~\cite{opal}, the branching ratio to these states
was estimated to
be $34\pm
7\%$.   From Ref.~\cite{fpesk}, we take the ratio of $D$ to $D^*$
production in
$B\to X_c\ell\nu$, for which several experimental measurements have been
combined consistently:
\begin{equation}
  {\Gamma(B\rightarrow D^* e\bar\nu_e)\over \Gamma(B\rightarrow D
  e\bar\nu_e)+\Gamma(B\rightarrow D^* e\bar\nu_e)}=0.65\pm 0.06\,.
\end{equation}
We estimate the minimum value for the
first moment of the invariant
mass spectrum by taking the $1\sigma$ limits of these experimental results.
Hence, we take a 27\% branching fraction to the $P$ wave states, and assume
that
the rest of the
branching fraction is saturated by the $D$ and $D^*$ in the ratio
0.41:0.59. The minimum value for the first moment
$\langle s_H-\bar m_D^2 \rangle$ is then
\begin{eqnarray}
   \langle s_H-\bar m_D^2\rangle_{\rm min.}&\simeq& 0.27\left[(2.450\,{\rm
   GeV})^2-(1.975\,{\rm GeV})^2\right]+0.43\left[(2.010\,{\rm
   GeV})^2-(1.975\,{\rm GeV})^2\right]\nonumber\\
   &&\mbox{} +0.30\left[(1.869\,{\rm
   GeV})^2-(1.975\,{\rm GeV})^2\right]\\
   &=&0.51\,{\rm GeV^2}\,. \nonumber
\end{eqnarray}
For the second moment, we will be conservative and neglect the small (and
positive) contribution of the ground state doublet.  We find
\begin{equation}\label{bound2}
   \langle (s_H-\bar m_D^2)^2\rangle_{\rm min.}\simeq
   0.27\times\left[(2.450\,{\rm
   GeV})^2-(1.975\,{\rm GeV})^2\right]^2=1.2\,{\rm GeV^4}.
\end{equation}
Solving Eq.~(\ref{massivetwo}) for the first moment, we find
\begin{equation}\label{lambarlim}
 \bar\Lambda>\left[0.41-1.41\,{\alpha_s\over\pi}-0.07\left({\lambda_1\over
 0.1\,{\rm GeV}^2}\right)\right]{\rm GeV}.
\end{equation}
The nonperturbative parameters $\bar\Lambda$ and $\lambda_1$ are
well-defined
only at a given order in perturbation theory~\cite{renorms}.  Our limits
apply
to these quantities defined at one loop.  We will use the coupling constant
$\alpha_s(m_b)=0.2$ in what follows.
Since $\lambda_1$ is closely related to minus the kinetic energy of the $b$
quark in the $B$ meson, it is expected to be negative.  Under this
assumption,
we obtain the lower bound
\begin{equation}
   \bar\Lambda>340\,{\rm MeV}.
\end{equation}
This limit corresponds to an upper bound on the $b$ quark pole mass of
$m_b^{\rm pole}<4.97\,{\rm GeV}$.
In Ref.~\cite{bsuvlam}, the stringent inequality
$\lambda_1\le-3\lambda_2\approx-0.35\,{\rm GeV}^2$ was proposed; in such a
case we would find the more restrictive bound
\begin{equation}
   \bar\Lambda>570\,{\rm GeV},
\end{equation}
corresponding to the upper limit $m_b^{\rm pole}<4.71\,{\rm GeV}$.

\begin{figure}
\epsfxsize=13cm
\hfil\epsfbox{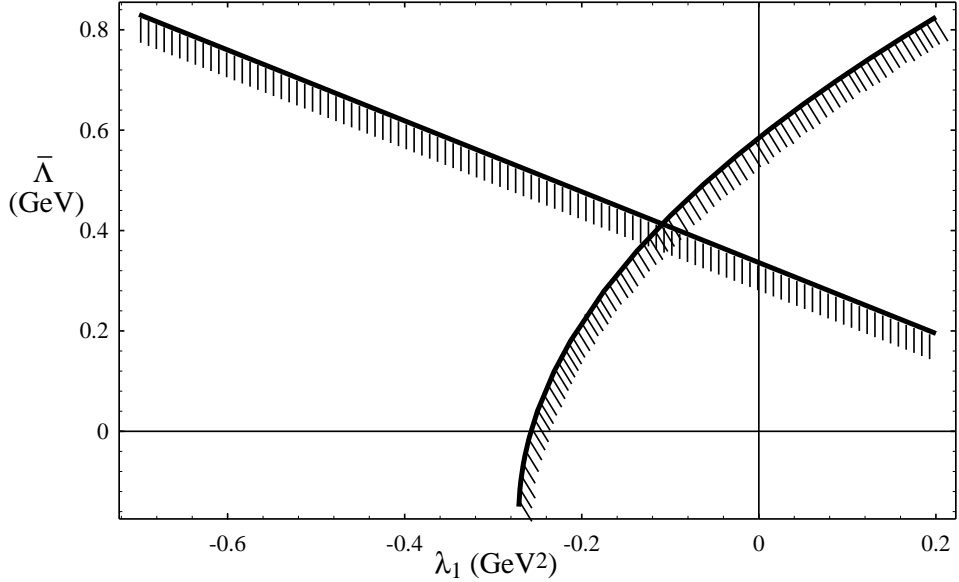}\hfill
\caption{Correlated one-loop limits on $\bar\Lambda$ and $\lambda_1$.  The
shaded region is ruled out by our analysis of the first two moments of
$(s_H-\bar m_D^2)$.}
\label{lam1lambar}
\end{figure}

If we also use the bound~(\ref{bound2}) on the second moment, we may relax
the
assumptions on $\lambda_1$ and obtain correlated limits on $\bar\Lambda$
and
$\lambda_1$.  These are plotted in Fig.~\ref{lam1lambar}.
By this method, we obtain the lower bound
\begin{equation}
   \bar\Lambda>410\,{\rm MeV}\,,
\end{equation}
independent of $\lambda_1$.  Where the bound on $\bar\Lambda$ is saturated,
$\lambda_1=-0.11\,{\rm GeV}^2$.  Our result implies the upper limit
$m_b^{\rm pole}<4.89\,{\rm GeV}$, {\it without\/} any assumption on
$\lambda_1$
being made.

This approach complements the recent proposal~\cite{KL} that $\bar\Lambda$
and
$\lambda_1$ be extracted from moments of the photon energy spectrum in the
rare
process $B\to X_s\gamma$.

\section{Higher Loops}

In order to apply our results consistently, it is important to know the
scale at which to evaluate
$\alpha_s(\mu)$ in the radiative corrections.  It has been shown recently
\cite{LSWpqcd} that the na\"\i ve choice $\mu=m_b$ significantly
underestimates
the size of
the two-loop effects.   In particular,  the prescription of Brodsky,
Lepage and
Mackenzie (BLM)~\cite{BLM} suggests that the relevant scale for the
radiative
corrections in
$B\rightarrow X_u e\bar\nu$ decay is $\mu\sim 0.07\, m_b$, when
expressed in terms of the $b$ quark pole mass, indicating that
two-loop effects are substantial.

It has also been stressed, however, that the BLM prescription may give a
misleadingly low scale when relating unphysical quantities
\cite{NUral,BBB2}.  In particular,
$\bar\Lambda$ is related to the pole mass of the heavy quark, which is not
an
observable, and in fact suffers from an inherent ambiguity in its
definition
\cite{renorms}.   In this
section, we show that although the BLM prescription indicates that
radiative corrections to the first two moments of the
invariant mass spectrum for semileptonic $b\to u$ decay are uncontrolled
when
expressed in terms of the HQET
parameter $\bar\Lambda$, they are well behaved when expressed in terms of
physical quantities.

The portion of the two loop correction to Eq.~(\ref{masslesstwo}) which is
proportional to the QCD evolution parameter $\beta_0$ may be determined
from
the one loop correction, calculated with a massive gluon in the final
state,
using the techniques of Ref.~\cite{SmithVol}.  Some of the
details of the computation are given in Appendix B; we find, for $\hat
m_q=0$,
\begin{eqnarray}\label{badseries}
  {1\over m_B^2}\langle s_H \rangle&=&
  {91\over 450}{\alpha_s(m_b)\over\pi}+\left({53\over
  180}\pi^2-{276043\over
  108000}\right)\beta_0\left({\alpha_s(m_b)\over\pi}\right)^2+{7\over 10}
  {\bar\Lambda\over m_B}+\dots\\
  &\simeq&0.20{\alpha_s(m_b)\over\pi} +
  3.15\left({\alpha_s(m_b)\over\pi}\right)^2
  + {7\over 10} {\bar\Lambda\over m_B}+\dots\nonumber \\
  &\simeq& 0.013+0.013+{7\over 10} {\bar\Lambda\over m_B}+\dots\,,\nonumber
\end{eqnarray}
where $\beta_0=11-2n_f/3$ and in the last line we have taken
$\alpha_s(m_b)\simeq 0.2$.
Clearly the perturbation expansion is poorly controlled.
In the BLM scale-setting prescription, the scale
$\mu_{\rm BLM}$  of the coupling is chosen such that the two-loop
contribution proportional to $\beta_0$ is absorbed into the one-loop
correction.  The poor convergence of the series is reflected
in the low BLM scale for this process:
\begin{equation}
  \mu_{\rm BLM}=m_b\exp\left[-2\left({53\over 180}\pi^2-{276043\over
  108000}\right)/{91\over 450}\right]\simeq 0.03 m_b\simeq 140\,{\rm
  MeV}\,.
\end{equation}
However, our expression for $\langle s_H\rangle$ is given in terms of the
unphysical parameter $\bar\Lambda$.  While this is perfectly acceptable as
an
intermediate step, since we
are ultimately interested only in relations between observable quantities,
it
has the effect of making the
perturbative expansion appear ill-behaved. Instead, let us define the
``decay mass" of
the $b$ quark, ${m_b^\Gamma}$, via the charmless semileptonic partial width
of the
$B$ meson,
\begin{equation} \Gamma(B\rightarrow X_u e \bar \nu_e)\equiv
  {G_F^2|V_{ub}|^2\over 192\pi^3} ({m_b^\Gamma})^5.
\end{equation}
The decay mass ${m_b^\Gamma}$ is a physical observable and is therefore
well-defined.
It is related to the pole mass via the expansion
\begin{equation}\label{massreln}
  {m_b^\Gamma}=m_b^{\rm pole}\left[1+\left(
  {5\over6}-{2\over15}\pi^2\right){\alpha_s(m_b)\over\pi}
  -(0.596\beta_0+c)\left({\alpha_s(m_b)\over\pi}\right)^2  + \dots +
  {\lambda_1-9\lambda_2\over 10 m_b^2}+ \dots\right]\,.
\end{equation}
The two-loop term proportional to $\beta_0$, which
one expects to dominate the two loop result, was calculated in
Ref.~\cite{LSWpqcd}.  The
constant $c$ has not been computed. Since $m_b^{\rm pole}$ is not
well-defined due to
renormalon effects, the perturbation series in Eq.~(\ref{massreln}) has a
renormalon ambiguity at ${\cal O}(1/m_b)$.

Defining a physical version of the parameter
$\bar\Lambda$,\footnote{Note that unlike $\bar\Lambda$, $m_Q-m_q^\Gamma$
is not universal for heavy
quarks, and differs in the $b$ and $c$ systems.  Since it explicitly
violates
heavy quark symmetry, it is not useful to reformulate HQET in terms
of this more physical quantity.}
\begin{equation}
 \bar\Lambda_b^\Gamma\equiv m_B-{m_b^\Gamma}\,,
\end{equation}
we have
\begin{equation}
  {\bar\Lambda\over m_B}={\bar\Lambda_b^\Gamma\over m_B}+\left(
  {5\over6}-{2\over15}\pi^2\right){\alpha_s(m_b)\over\pi}
  -(0.596\beta_0+c)\left(
  {\alpha_s(m_b)\over\pi}\right)^2+{\cal O}(\alpha_s^3,1/m_b^2)\,,
\end{equation}
and Eq.~(\ref{badseries}) becomes
\begin{eqnarray}\label{notbadseries}
  {1\over m_B^2}\langle \hat s_H \rangle
  &\simeq&(0.202-0.337){\alpha_s(m_b)\over\pi} +
  (3.151-3.752)\left({\alpha_s(m_b)\over\pi}\right)^2 + {7\over 10}
  {\bar\Lambda_b^\Gamma\over m_B} \nonumber \\
  &\simeq& -0.135{\alpha_s(m_b)\over \pi}-0.601
  \left({\alpha_s(m_b)\over\pi}\right)^2+{7\over 10}
{\bar\Lambda_b^\Gamma\over m_B}
  \nonumber \\
  &\simeq&-0.0086-0.0024+{7\over 10} {\bar\Lambda_b^\Gamma\over m_B}\,.
\end{eqnarray}
The perturbation expansion clearly has improved dramatically.  The
corresponding
BLM scale is now
\begin{equation}
  \mu_{\rm BLM}=m_b\exp\left[-(2/9)0.601/0.135\right]\simeq 0.37\,m_b\,,
\end{equation}
which is significantly greater than before.

It is interesting to note that the cancellation we observe in
Eq.~(\ref{notbadseries}) persists
at higher orders in the bubble sum.  Using the techniques of
Ref.~\cite{BBB}
we can calculate the $n$ loop bubble graph, from which we may extract the
coefficient of $\beta_0^n\alpha_s^{n+1}$ in the perturbative expansion for
$\langle\hat s_0\rangle$.  Although there is no reason to believe that this
is the dominant contribution at this order, since there is no
$\beta_0\rightarrow \infty$
limit of QCD in
which the quark and gluon bubble graphs dominate, it does give one class of
contributions to the $n$
loop graphs which displays a factorial divergence at large orders in
perturbation theory.

Using the techniques of Ref.~\cite{BBB}, the perturbation series in
Eq.~(\ref{badseries}) continues as
\begin{eqnarray}\label{morebadseries}
  {1\over m_B^2}\langle \hat s_H \rangle&=&
  0.202{\alpha_s(m_b)\over\pi}+3.151\left({\alpha_s(m_b)\over\pi}\right)^2
  +51.91\left({\alpha_s(m_b)\over\pi}\right)^3\\
  &&+940.52\left({\alpha_s(m_b)\over\pi}\right)^4
  +19347.5\left({\alpha_s(m_b)\over\pi}\right)^5+{7\over10}
  {\bar\Lambda\over m_B}+\dots\,, \nonumber
\end{eqnarray}
and using the results of Ref.~\cite{BBB2} for the higher order relation
between
$m_b$ and ${m_b^\Gamma}$, Eq.~(\ref{notbadseries}) continues as
\begin{eqnarray}\label{morenotbadseries}
  {1\over m_B^2}\langle \hat s_H \rangle
  &=&(0.202-0.337){\alpha_s(m_b)\over\pi}
  +(3.151-3.752)\left({\alpha_s(m_b)\over\pi}\right)^2\nonumber \\
  &&+(51.91-50.37)\left({\alpha_s(m_b)\over\pi}\right)^3
  +(940.52-782.42)\left({\alpha_s(m_b)\over\pi}\right)^4\nonumber \\
  &&+(19347.5-14424.2)\left({\alpha_s(m_b)\over\pi}\right)^5+ {7\over 10}
  {\bar\Lambda_b^\Gamma\over m_B} +\dots \nonumber \\
  &=&-0.135{\alpha_s(m_b)\over\pi}
  -0.601\left({\alpha_s(m_b)\over\pi}\right)^2
  +1.56\left({\alpha_s(m_b)\over\pi}\right)^3\\
  &&+148.1\left({\alpha_s(m_b)\over\pi}\right)^4
  +4923.\left({\alpha_s(m_b)\over\pi}\right)^5
  +{7\over 10} {\bar\Lambda_b^\Gamma\over m_B}+\dots \nonumber \\
  &\simeq&-0.0086-0.0024+0.0004+0.0026+0.0051
  +{7\over 10} {\bar\Lambda_b^\Gamma\over m_B}+\dots \nonumber
\end{eqnarray}
Note that even at higher orders there is significant cancellation between
the
two series.  This is similar to the behaviour observed in a different
context
in Ref.~\cite{BBB2}.  The remaining bad behaviour presumably
reflects the presence of
unphysical parameters (such as $\bar\Lambda^2$ and $\lambda_1$) at higher
orders
in the operator product expansion.
Assuming the series is asymptotic, the size of the smallest term in the
expansion gives a measure of the uncertainty in the sum of
the series.

We do not find a similar cancellation for the second moment of $s_H$.  For
$\hat
m_q=0$ and to order $1/m_b$, we find
\begin{equation}
   {1\over m_B^4}\langle s_H^2\rangle=\langle \hat s_0^2
   \rangle+4{\bar\Lambda\over m_B}
   \left(\langle \hat E_0\hat s_0 \rangle - \langle \hat s_0^2
   \rangle\right)+\dots\,.
\end{equation}
Since $\langle \hat E_0\hat s_0 \rangle$ and $\langle \hat s_0^2 \rangle$
are both order $\alpha_s$, there is no $\beta_0\alpha_s^2$ term introduced
by expressing $\langle s_H^2\rangle$ in terms of $\bar\Lambda_b^\Gamma$.
However, the na\"\i ve counting of powers of $\beta_0$ does not work here,
because
$\langle \hat s_0^2\rangle\ll \langle \hat E_0 \hat s_0\rangle, \langle
\hat
s_0 \rangle$.  Instead, the ${\cal O}(\beta_0\alpha_s^2)$ correction to
$\langle\hat s_0^2\rangle$ is the same order as the ${\cal O}(\alpha_s^2)$
term
introduced by expressing $\langle s_H^2\rangle$ in terms of
$\bar\Lambda_b^\Gamma$.
Using the $\beta_0\alpha_s^2$ term as an estimate of the full two loop
correction to
$\langle s_0^2\rangle$ alone, we find, using the same technique as before,
\begin{equation}\label{s0series}
   \langle s_0^2\rangle={5\over 324}{\alpha_s(m_b)\over\pi}+
   \left({277\over 648}\pi^2-{25511\over6075}\right)\beta_0
   \left({\alpha_s(m_b)\over\pi}\right)^2+\dots\,,
\end{equation}
and so
\begin{eqnarray}
   {1\over m_B^4}\langle s_H^2\rangle&=&0.015{\alpha_s(m_b)\over\pi}+
   0.0196\beta_0\left({\alpha_s(m_b)\over\pi}\right)^2
   +0.298\left({\bar\Lambda_b^\Gamma\over m_B}-0.48
   {\alpha_s(m_b)\over\pi}\right){\alpha_s(m_b)\over\pi}\nonumber \\
   &&\qquad\mbox{}+0.533\left({\bar\Lambda_b^\Gamma\over m_B}-0.48
   {\alpha_s(m_b)\over\pi}\right)^2+\ldots \\
   &=&0.015{\alpha_s(m_b)\over\pi}+0.156
   \left({\alpha_s(m_b)\over\pi}\right)^2
   -0.214{\bar\Lambda_b^\Gamma\over m_B}{\alpha_s(m_b)\over\pi}+
   \dots \nonumber\\
   &\simeq& 9.5\times 10^{-4}+6.3\times 10^{-4}
   -0.214{\bar\Lambda_b^\Gamma\over
m_B}{\alpha_s(m_b)\over\pi}+\dots\,.\nonumber
\end{eqnarray}
The contribution of the $\beta_0\alpha_s^2$ term to this expression is
$7.1\times10^{-4}$, and the new ${\cal O}(\alpha_s^2)$ terms, while of the
same
order as this one, largely cancel against each other.  Since the
convergence of
the perturbation series for $\langle s_H^2\rangle$ still appears to be
poor, we
may also expect the limits on $\bar\Lambda$ and $\lambda_1$ which we
obtained
from $\langle (s_H-\bar m_D^2)^2\rangle$ to be more sensitive to higher
order
perturbative corrections than those obtained from $\langle s_H-\bar
m_D^2\rangle$.

Finally, note that the appearance of $\bar\Lambda$ in $\langle
s_H^2\rangle$ is
suppressed by a factor of $\alpha_s$, as is its associated renormalon at
${\cal
O}(1/m_b)$.  Since renormalon ambiguities must cancel in relations between
physical quantities, this means that the large $\beta_0\alpha_s^2$ term in
$\langle s_H^2\rangle$ does not correspond to a ${\cal O}(1/m_b)$
renormalon
ambiguity in the perturbation series (\ref{s0series}).

\section{Summary and Conclusions}

We have used the operator product expansion and the heavy quark limit to
compute the  hadronic energy and invariant mass spectra in semileptonic
heavy
meson decays.  Our expressions are complete up to order $\alpha_s$ in
perturbation
theory, and up to order $\alpha_s/m_b$ and $1/m_b^2$ in the heavy quark
expansion.
The effects of finite
final state quark masses have been taken into account, so it is possible to
apply
our results to the important decay $b\to c\,\ell\nu$.

Our analysis provides a test of the applicability of the OPE to these
decays, and
of the crucial underlying concept of global duality.  Only appropriately
weighted integrals
of the theoretical spectra may be compared meaningfully  with experiment,
and we focus
on the leading moments.  As an initial application, we used the recent
measurement
of the $B$ branching fraction to excited $D$ mesons to put bounds
on the nonperturbative parameters $\bar\Lambda$ and $\lambda_1$.  We found
$\bar\Lambda>410\,$MeV, which led to a constraint on the $b$ quark pole
mass,
$m_b^{\rm pole}<4.89\,{\rm GeV}$.
More stringent tests will have to await the availability of more precise
data.
The success or failure of our
predictions will determine the confidence with which one will trust these
theoretical
techniques in the extraction of CKM matrix elements from semileptonic
bottom
and charm decays.

We also investigated the behaviour of the perturbation series at higher
order
in
$\alpha_s$, to gain insight into the trustworthiness of the lowest order
calculation and
the choice of renormalization scale $\mu$.  We found that when written in
terms of the
unphysical quantity $\bar\Lambda$, the perturbation series for $\langle
s_H\rangle$
seems to be quite badly behaved, with a BLM scale
$\mu_{{\rm BLM}}$ too low to be meaningful.  However, when we define a more
physical
``decay mass'' $m_b^\Gamma$, and through it a physical
$\bar\Lambda_b^\Gamma$, the
perturbation series improves dramatically.  The cancellations which we find
persist to
higher order in
$\alpha_s$, at least when one includes the leading powers of $\beta_0$.

We have focused on the application to $B$ decays; however, the BLM analysis
suggests that the perturbative corrections to $\langle s_H\rangle$ are
under
control for $D$ decays as well.  The extension of our results to charm is
straightforward.

\acknowledgements
It is a pleasure to acknowledge helpful conversations with Mark Wise and
Lincoln Wolfenstein.  This work was initiated at the Weak Interactions
Workshop at the
Institute for Theoretical Physics in Santa Barbara, and we thank the
organizers
for their gracious and efficient hospitality.  We are
equally grateful to the High  Energy  Theory group at the University of
California, San Diego, for their generous support.  M.L.~and M.S.~also
thank the
Institute for Nuclear Theory at the University of Washington.  This
work was supported by the United States Department of Energy  under Grant
Nos.~DE-FG03-90ER40546 and DE-FG02-91ER40682, by the United  States
National
Science Foundation under Grant No.~PHY-9404057, and by the Natural
Sciences and Engineering Research Council of Canada.  A.F.~also
acknowledges
the United States National Science Foundation for National Young
Investigator Award
No.~PHY-9457916, and the United States Department of Energy for
Outstanding Junior Investigator Award No.~DE-FG02-94ER40869.

\appendix

\section{The parton level hadronic energy spectrum}

In this appendix we discuss the corrections to the parton level hadronic
energy spectrum, ${\rm d}\Gamma_0/{\rm d}\hat E_0$.  Both the
perturbative and the power corrections are somewhat unwieldy;
we present them here for completeness.

The power correction may be computed by integrating the doubly
differential spectrum~(\ref{diffspect}) over $\hat s_0$.  The integral
will be nonzero only if $\hat E_0\le{1\over2}(1+\hat m_q^2)$, because, as
discussed in Section II, only in this case does the integration region
overlap with the one-particle pole at $\hat s_0=\hat m_q^2$.  This is a
reflection of the fact that the maximum energy a single quark can carry
away from the decay is
${1\over2}(1+\hat m_q^2)$.  In the presence of additional  strongly
interacting particles such as gluons, the total hadronic energy $\hat E_0$
can exceed ${1\over2}(1+\hat m_q^2)$.

However, the initial motion of the $b$ quark inside the $B$ meson   can
produce fluctuations of the maximum allowed final quark energy above
${1\over2}(1+\hat m_q^2)$.  These fluctuations appear in the differential
rate as   singular functions $\delta(\hat E_0-{1\over2}(1+\hat m_q^2))$
and $\delta'(\hat E_0-{1\over2}(1+\hat m_q^2))$, which are resummed into a
smooth function extending beyond the  parton model endpoint. For
a more detailed discussion of this subject see
Refs.~\cite{MW,Neubert,FJMW}.

Including the leading power corrections, then, the  expression
for the hadronic energy spectrum is given by\footnote{We do not agree
with the expression presented in Ref.~\cite{BZ}.}
\begin{eqnarray}
  {1\over\Gamma_0}{{\rm d}\Gamma\over{\rm d}\hat E_0}
  &=&
  16\sqrt{\hat E_0^2-\hat m_q^2}\left[3\hat E_0-4\hat E_0^2-2\hat m_q^2
   +3\hat E_0\hat m_q^2\right]\nonumber\\
  &&+{16\over\sqrt{\hat E_0^2-\hat m_q^2}}\left[{\lambda_1\over2m_b^2}
  \left(-6\hat E_0^2+12\hat E_0^3+{20\over3}\hat E_0^4+3\hat m_q^2
  -6\hat E_0\hat m_q^2
  -{52\over3}\hat E_0^2\hat m_q^2+{23\over3}\hat m_q^4\right)\right.
  \nonumber\\
  &&\qquad\left.+{\lambda_2\over 2m_b^2}\left(
  -3\hat E_0-6\hat E_0^2+36\hat E_0^3+20\hat E_0^4+3\hat m_q^2-21\hat E_0
   \hat m_q^2
  -52\hat E_0^2\hat m_q^2+23\hat m_q^4\right)\right] \nonumber\\
  && + (1-\hat m_q^2)^3 \,\left[
  \displaystyle{\lambda_1\over 3m_b^2} (5-\hat m_q^2)
  -\displaystyle{\lambda_2\over m_b^2}  (1-5 \hat m_q^2) \right]\
  \delta\left(\hat E_0-{1\over2}(1+\hat m_q^2)\right)\,  \nonumber \\
  &&+ \displaystyle{\lambda_1\over 6 m_b^2}
 (1-\hat m_q^2)^5 \ \delta'\left(\hat E_0-{1\over2}(1+\hat
  m_q^2)\right)+{\cal O}(\alpha_s,1/m_b^3)\,.
\end{eqnarray}
Integrating this expression with respect to $\hat E_0$, we find  the
power corrections~(\ref{e0np}) to the moments
$\langle\hat E_0\rangle$ and $\langle\hat E_0^2\rangle$.

\begin{figure}
\epsfxsize=12cm
\hfil\epsfbox{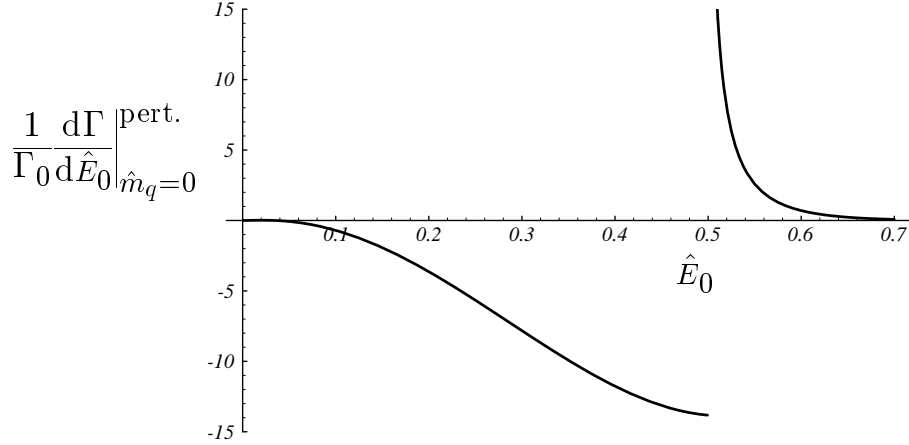}\hfill
\caption{The order $\alpha_s$ contribution to the differential
energy spectrum
$(1/\Gamma_0){\rm d}\Gamma/{\rm d}\hat E_0$, for $\hat m_q=0$, in
units of $\alpha_s/\pi$.  In the region $\hat E_0>{1\over2}$,
this is the leading nonzero
contribution.  The logarithmic
divergence at $\hat E_0={1\over2}$ is integrable.}
\label{tradplot}
\end{figure}

The expression for the perturbative correction to the hadronic energy
spectrum is even more cumbersome.  For the complete spectrum at finite
$\hat m_q$, we refer the reader to Ref.~\cite{CJK}.  As an illustration we
present
here the perturbative corrections at
$\hat m_q=0$, separately for $\hat E_0<{1\over2}$  and
$\hat E_0> {1\over2}$:
\begin{eqnarray}
     \left.{1\over\Gamma_0}{{\rm d}\Gamma\over{\rm d}\hat E_0}\right
     \vert_{0<\hat E_0<{1\over2}}^{\rm pert.}
     &=&{\alpha_s\over\pi}\,
     E_0^2\bigg[36-{32\over3}\pi^2-{496\over9}\hat E_0
     +{128\over9}\pi^2\hat E_0+{52\over3}\hat E_0^2
     -{112\over45}\hat E_0^3+{64\over135}\hat E_0^4\nonumber\\
     &&\qquad\qquad-24\ln (2\hat E_0)+{64\over3}\hat E_0\ln(2\hat E_0)
     +16\ln (2\hat E_0)\ln(1-2\hat E_0)\\
     &&\qquad\qquad-{64\over3}\hat E_0\ln (2\hat E_0)\ln(1-2\hat E_0)
     +16{\rm Li}_2(2\hat E_0)
     -{64\over3}\hat E_0{\rm Li}_2(2\hat E_0)\bigg]\,,\nonumber\\
     \left.{1\over\Gamma_0}{{\rm d}\Gamma\over{\rm d}\hat E_0}\right
     \vert_{{1\over2}<\hat E_0<1}^{\rm pert.}
     &=&{\alpha_s\over\pi}\Bigg[{208\over45}+{1058\over45}\hat E_0
     -{646\over9}\hat E_0^2+{1592\over27}\hat E_0^3
     -{52\over3}\hat E_0^4+{112\over45}\hat E_0^5
     -{64\over135}\hat E_0^6\nonumber\\
     &&\qquad+{5\over9}\ln(2\hat E_0-1)+{8\over3}\hat E_0\ln(2\hat E_0-1)
     +{16\over3}\hat E_0^2\ln(2\hat E_0-1)\\
     &&\qquad-{64\over9}\hat E_0^3\ln(2\hat E_0-1)+8\hat E_0^2\ln^2
     (2\hat E_0-1)
     -{32\over3}\hat E_0^3\ln^2(2\hat E_0-1)\nonumber\\
     &&\qquad-16\hat E_0^2\ln (2\hat E_0)\ln(2\hat E_0-1)
     +{64\over3}\hat E_0^3\ln (2\hat E_0)\ln(2\hat E_0-1)\nonumber\\
     &&\qquad-16\hat E_0^2{\rm Li}_2\left({1\over 2\hat E_0}\right)
     +{64\over3}\hat E_0^3{\rm Li}_2\left({1\over 2\hat E_0}\right)
     +16\hat E_0^2{\rm Li}_2\left({2\hat E_0-1\over 2\hat E_0}\right)
     \nonumber\\&&\qquad-{64\over3}\hat E_0^3{\rm Li}_2
     \left({2\hat E_0-1\over2\hat E_0}\right)\Bigg]\,.\nonumber
\end{eqnarray}
This spectrum is shown in Fig.~\ref{tradplot}.  The logarithmic
divergence as $\hat E_0\to{1\over2}$ from above is integrable.  The region
$\hat E_0>\case1/2$ receives contributions only from brehmsstrahlung
graphs.
Note that the spectrum falls extremely rapidly with increasing $\hat E_0$.

The radiative corrections $A_m(\hat m_q)$ to the moments $\langle\hat
E_0^m\rangle$ may be obtained by integrating the full expressions found in
Ref.~\cite{CJK}.  We find
\begin{eqnarray}\label{ratecorrn}
  A_0(\hat m_q)=&&{25\over 6}-{2\over 3}\pi^2-{478\over
  9}\hat m_q^2+{64\over 3}\pi^2(1+\hat m_q^2)\hat m_q^3-{32\over
  3}\pi^2\hat m_q^4
  +{478\over 9} \hat m_q^6\nonumber\\
  &&-\left({25\over 6}+{2\over 3}\pi^2\right)
  \hat m_q^8-{2\over 3}(36+\hat m_q^4)\hat m_q^4\ln^2\hat m_q^2\nonumber\\
  &&+\left(-{40\over 3}\hat m_q^2+{256\over 3}(1+\hat m_q^2)\ln(1+\hat m_q)
  \hat m_q^3-60\hat m_q^4+{8\over 9}\hat m_q^6-
  {34\over 9}\hat m_q^8\right)\ln\hat m_q^2\nonumber \\
  &&+\left(-{34\over 9}+{128\over 9}\hat m_q^2-{128\over 9}\hat
m_q^6+{34\over
  9}\hat m_q^8\right)\ln(1-\hat m_q^2)\\
  &&+\left({8\over3}-{128\over 3}(1+\hat m_q^2)\hat m_q^3+80\hat
m_q^4+{8\over
  3}\hat m_q^{8}\right)\ln\hat m_q^2\ln(1-\hat m_q^2)\nonumber \\
  &&+\left(4+{128\over 3}(1+\hat m_q^2)\hat m_q^3+64\hat m_q^4+4 \hat
  m_q^8\right){\rm Li}_2(\hat m_q^2) \nonumber\\
  &&-{512\over 3}\hat m_q^3(1+\hat m_q^2) {\rm Li}_2(\hat m_q)\,,\nonumber
\end{eqnarray}
\begin{eqnarray}
  A_1(\hat m_q)=&&{1381\over 900}-{7\over 30}\pi^2-\left({3133\over
  900}+{5\over 18}\pi^2\right)\hat m_q^2+\left({99329\over 1350}+{16\over
  3}\pi^2\right)
  \hat m_q^4-{1408\over 45}\pi^2 \hat m_q^5 \nonumber \\
  &&-\left({100729\over1350}-{16\over 3}\pi^2\right)\hat
m_q^6+\left({4933\over
  900}-{5\over 18}\pi^2\right)\hat m_q^8
  -\left({6743\over 2700}+{7\over 30}\pi^2\right)\hat m_q^{10}\nonumber \\
  &&+\left(6\hat m_q^4+{34\over 3}\hat m_q^6-{5\over 18}\hat m_q^8-{7\over
30}
  \hat m_q^{10}\right)\ln^2\hat m_q^2 \\
  &&+\left(-{47\over 30}\hat m_q^2+{1651\over 45}\hat m_q^4-{5632\over
  45}\ln(1+\hat m_q)\hat m_q^5+{1391\over 45}\hat m_q^6+
  {121\over 135}\hat m_q^8-{409\over 450}\hat m_q^{10}\right)\ln\hat m_q^2
  \nonumber \\
  &&+\left(-{61\over 50}+{97\over 54}\hat m_q^2-4\hat m_q^4+4\hat
  m_q^6-{97\over 54}
  \hat m_q^8+{61\over 50}\hat m_q^{10}\right)\ln(1-\hat m_q^2)\nonumber \\
  &&+\left({14\over 15}+{10\over 9}\hat m_q^2-{100\over 3}\hat
m_q^4+{2816\over
  45}\hat m_q^5
  -{100\over 3}\hat m_q^6+{10\over 9}\hat m_q^8+{14\over 15}\hat
  m_q^{10}\right)
  \ln\hat m_q^2\ln(1-\hat m_q^2)\nonumber \\
  &&+\left({7\over 5}+{5\over 3}\hat m_q^2-32\hat m_q^4-{2816\over 45}\hat
  m_q^5-32\hat m_q^6+{5\over 3}\hat m_q^8+{7\over 5}\hat m_q^{10}\right)
  {\rm Li}_2(\hat m_q^2) \nonumber\\
  &&+{11264\over 45}\hat m_q^5\, {\rm Li}_2(\hat m_q)\,,\nonumber
\end{eqnarray}
and
\begin{eqnarray}
  A_2(\hat m_q)=&&{2257\over 3600}-{4\over 45}\pi^2+\left({2929\over
  5400}-{1\over 5}\pi^2\right)\hat m_q^2-{324727\over 10800}
  \hat m_q^4+{64\over 5}\pi^2(1+\hat m_q^2) \hat m_q^5\nonumber \\
  &&+\left({173\over162}-{208\over
  27}\pi^2\right)\hat m_q^6+{304877\over10800}\hat m_q^8
  +\left({1297\over 1800}-{1\over 5}\pi^2\right)\hat m_q^{10}
  \nonumber \\
  &&-\left({36283\over 32400}+{4\over 45}\pi^2\right)\hat m_q^{12}
  -\left({116\over 9}\hat m_q^6+{1\over 5}\hat m_q^{10}+{4\over
  45}\hat m_q^{12}\right)\ln^2\hat m_q^2 \\
  &&+\left(-{2\over 45}\hat m_q^2-{131\over 20}\hat m_q^4+{256\over
  5}(1+\hat m_q^2)\ln(1+\hat m_q)\hat m_q^5-{5467\over 135}\hat m_q^6-
  {829\over 180}\hat m_q^8\right.\nonumber \\
  &&\left.\qquad+{23\over 450}\hat m_q^{10}-{173\over
  675}\hat m_q^{12}\right)\ln\hat m_q^2\nonumber \\
  &&+\left(-{298\over 675}+{1\over 25}\hat m_q^2+2\hat m_q^4-2\hat
  m_q^8-{1\over 25}
  \hat m_q^{10}+{298\over 675}\hat m_q^{12}\right)\ln(1-\hat
m_q^2)\nonumber \\
  &&+\left({16\over 45}+{4\over 5}\hat m_q^2-{128\over
  5}(1+\hat m_q^2)\hat m_q^5+{440\over9}\hat m_q^6
  +{4\over 5}\hat m_q^{10}+{16\over 45}\hat m_q^{12}\right)
  \ln\hat m_q^2\ln(1-\hat m_q^2)\nonumber \\
  &&+\left({8\over 15}+{6\over 5}\hat m_q^2+{128\over
  5}(1+\hat m_q^2)\hat m_q^5+{416\over9}\hat m_q^6+
  {6\over 5}\hat m_q^{10}+{8\over 15}\hat m_q^{12}\right)
  {\rm Li}_2(\hat m_q^2) \nonumber\\
  &&-{512\over5}(1+\hat m_q^2)\hat m_q^5\, {\rm Li}_2(\hat m_q)\,.
\nonumber
\end{eqnarray}
The correction $A_0(\hat m_q)$ to the total rate is equivalent to the
result
presented in Ref.~\cite{Nir}.

\section{Bubble Graphs}

The $n$-loop bubble graph contribution to moments of $\hat s_0$ may
be calculated from the one loop graph evaluated with a finite gluon mass
\cite{SmithVol,BBB}.  In this appendix, we briefly outline this calculation
using the methods of Ref.~\cite{BBB}.
Only the bremmstrahlung graphs in Fig.~\ref{graphs} contribute to
the moments of $\hat s_0$ for $n\geq 1$.   We consider the expansion
\begin{equation}
  {{\rm d}\Gamma\over{\rm d}\hat s_0}=\sum_{j=0}^\infty d_j(\hat s_0)
  \beta_0^j\left({\alpha_s\over\pi}\right)^{j+1}+\dots\,,
\end{equation}
where $\beta_0=11-2n_f/3$ and
the ellipses denote terms which have fewer powers of
$\beta_0\alpha_s$ and hence are not obtainable from the bubble graphs.
Note
that these are not suppressed terms in any limit of QCD, although they may
be
numerically small.  The $n^{\rm th}$ moment of $\hat s_0$ then has the
expansion
\begin{equation}
  {\cal{M}}^{(n,0)}=\sum_{j=0}^\infty
  m^{(n)}_j\,\beta_0^j\left({\alpha_s\over\pi}\right)^{j+1}+\ldots\,,
\end{equation}
where
\begin{equation}
  m^{(n)}_j=\int_0^1{\rm d}\hat s_0\, \hat s_0^n\,d_j(\hat s_0)\,.
\end{equation}

Define $d_0(\hat s_0,\hat\lambda^2)$ and $m^{(n)}_0(\hat\lambda^2)$ to be
the
one-loop corrections calculated with a finite gluon mass $\lambda$, and
$\hat\lambda\equiv\lambda/m_b$.  Then
\begin{equation}
  m^{(n)}_0(\hat\lambda^2)=\int_{\hat\lambda^2}^1{\rm d}\hat
s_0^n\,d_0(\hat
  s_0,\hat\lambda^2)\,,
\end{equation}
and we have \cite{BBB}
\begin{equation}\label{nloopa}
  d_j(\hat s_0)=\left.-{1\over 4^j}{{\rm d}^j\over
  {\rm d}u^j}\right\vert_{u=0}{\sin(\pi u)\over \pi u}
  \int_0^{\hat s_0}\left(\hat\lambda^2 e^C\right)^{-u}\,{{\rm d}\over
  {\rm d}\hat\lambda^2}d_0(\hat s_0,\hat\lambda^2)\,,
\end{equation}
where $C$ is a scheme-dependent constant.  In the $V$ scheme\cite{BLM},
$C=0$, while in the $\overline{{\rm MS}}$ scheme, $C=-5/3$.
Eqn.~(\ref{nloopa}) may be written in the form
\begin{equation}\label{nloop}
  d_j(\hat s_0)=\sum_{k=0}^j\,c_k\,J_k(\hat s_0)\,,
\end{equation}
where $J_k$ is defined by
\begin{equation}
  J_k(\hat s_0)\equiv\int_0^{\hat s_0}
  {\rm d}\hat\lambda^2\,\ln^k(\hat\lambda^2)\,{{\rm d}\over
  {\rm d}\hat\lambda^2}d_0(\hat s_0,\hat\lambda^2)\,.
\end{equation}.

Taking moments of both sides of Eq.~(\ref{nloop}), we have
\begin{eqnarray}\label{finalloop}
  m^{(n)}_j&=&\sum_{k=0}^j\, c_k\,\int_0^1{\rm d}\hat s_0\,\hat
  s_0^n\int_0^{\hat s_0}
  {\rm d}\hat\lambda^2\,\ln^k(\hat\lambda^2)\,{{\rm d}\over
  {\rm d}\hat\lambda^2}d_0(\hat s_0,\hat\lambda^2) \nonumber \\
  &=&\sum_{k=0}^j\, c_k\,\int_0^1
  {\rm d}\hat\lambda^2\,\int_{\hat\lambda^2}^1{\rm d}\hat s_0\,\hat s_0^n
  \ln^k(\hat\lambda^2)\,{{\rm d}\over
  {\rm d}\hat\lambda^2}d_0(\hat s_0,\hat\lambda^2) \\
  &=&\sum_{k=0}^j\, c_k\,\int_0^1 {\rm d}\hat\lambda^2\,
  \ln^k(\hat\lambda^2)\,{{\rm d}\over
  {\rm d}\hat\lambda^2}m_0^{(n)}(\hat\lambda^2)\,, \nonumber
\end{eqnarray}
where we have used the fact that $d_0(\hat\lambda^2,\hat\lambda^2)=0$ to
move
the $\hat s_0$ integral to the right of the $\hat\lambda^2$ derivative.

It is straightforward to derive analytic expressions for the moments
$m_0^{(n)}(\hat\lambda^2)$ from the graphs in Fig.~\ref{graphs};  however
the
resulting formulas are lengthy and we will not reproduce them here.
For $j=1$ the integrals in Eq.~(\ref{finalloop}) may be performed
analytically,
giving the  ${\cal O}(\alpha^2\beta_0)$ correction to ${\cal M}^{(n,0)}$,
while
for $j>1$ we performed the integrals numerically to obtain the contribution
from higher loops in the bubble sum.

\end{document}